# Splitting one ventilator for multiple patients – a technical assessment

T. Martinsen[1], C. Tronstad[1], M. Olsen[1], L.A. Rosseland[2], F.J. Pettersen[1], R.J. Strand-Amundsen[1], Ø.G. Martinsen[3,1], J.O. Høgetveit [1,3], H. Kalvøy[1]

1. Department of Clinical and Biomedical Engineering, Oslo University Hospital, Oslo, Norway
2. Division of Emergencies and Critical Care, Oslo University Hospital, Oslo, Norway
3. Department of Physics, University of Oslo, Oslo, Norway.

## Abstract

Due to the recent coronavirus outbreak, many efforts and innovative solutions have surfaced to deal with the possible shortage of ventilators upon catastrophic surges of patients. One solution involves splitting one ventilator to treat multiple patients and is in principle easy to implement, but there are obvious risks, and little is known on how the technique would work on patients with ARDS from Covid-19. Previous studies have shown that multiple test lungs of equal characteristics can be successfully ventilated from one machine, but that large variations in tidal volume delivery occurs when lungs with different compliance are connected. In contribution to the discussion of the feasibility of the technique, a technical assessment was done including experiments expanding on the previous studies using two types of test lungs, different ventilator settings and test lung characteristics. Using two test lungs connected to a ventilator, the tidal volumes and pressures into both lungs were measured for different combinations of lung compliance, airway resistances, modes of ventilation, inspiratory and end-expiratory pressure levels. We found discrepancies in delivered tidal volumes for paired test lungs proportional with compliance differences, little influence from differences in airway resistances, and that changes in compliance of only one test lung would also change the tidal volume delivered to the other test lung when in volume controlled mode. For one of the test lung types, we also found that higher PEEP settings could strongly influence the tidal volume balance between the test lungs. From this study and from a technical point of view, we were not able to identify reliable settings, adjustments or any simple measures to overcome the hazards of this simple technique, and a more advanced solution is indicated for mitigating risks.

## Introduction

During catastrophes with a surge of patients needing ventilatory support exceeding the number of available ventilators, a solution has been proposed for connecting multiple patients to the same ventilator. Neyman and Babcock Irvin introduced this concept in 2006, which is described below. The solution is very simple to implement, requiring only three T-tubes, adapters and extra tubes for each patient. With the recent coronavirus pandemic, interest in the solution has quickly spread globally, but little is known on how the solution works in reality, in particular for the treatment of ARDS from the coronavirus.

## Description of the solution

The Neyman and Babcock Irvin solution involves connecting four patients to one ventilator by four sets of tubes connected to the ventilator using two two-way splitters as shown in figure 1, one for the inspiratory and another for the expiratory limb. The splitter is made of three T-tubes and connection adapters connected as shown in figure 1b. Other types of splitters such as Y-connectors should work equally.

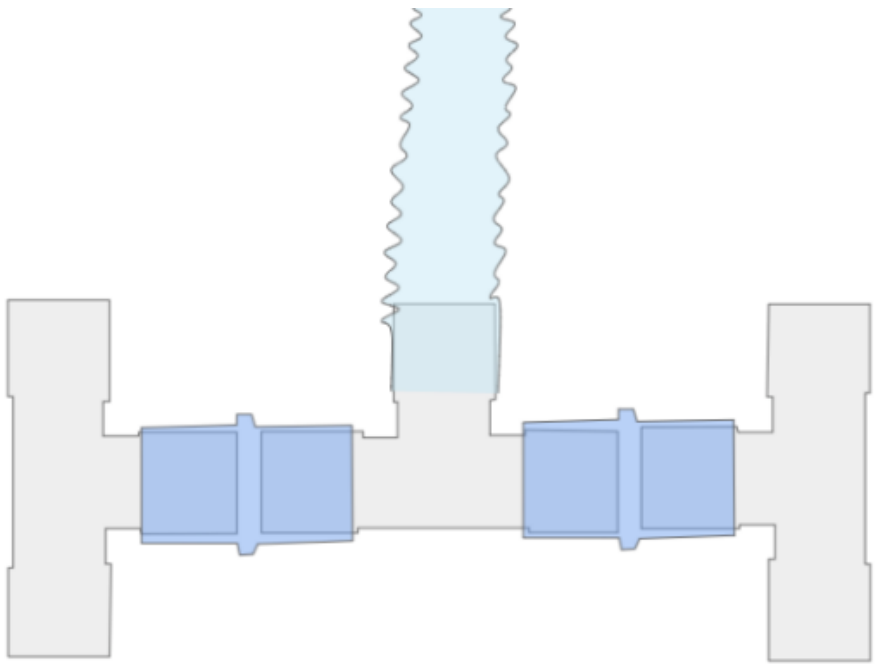

Figure 1. Example of a connection of three T-tubes and adapters to make a splitter for four patients connected to the inspiratory and expiratory limb of one ventilator.

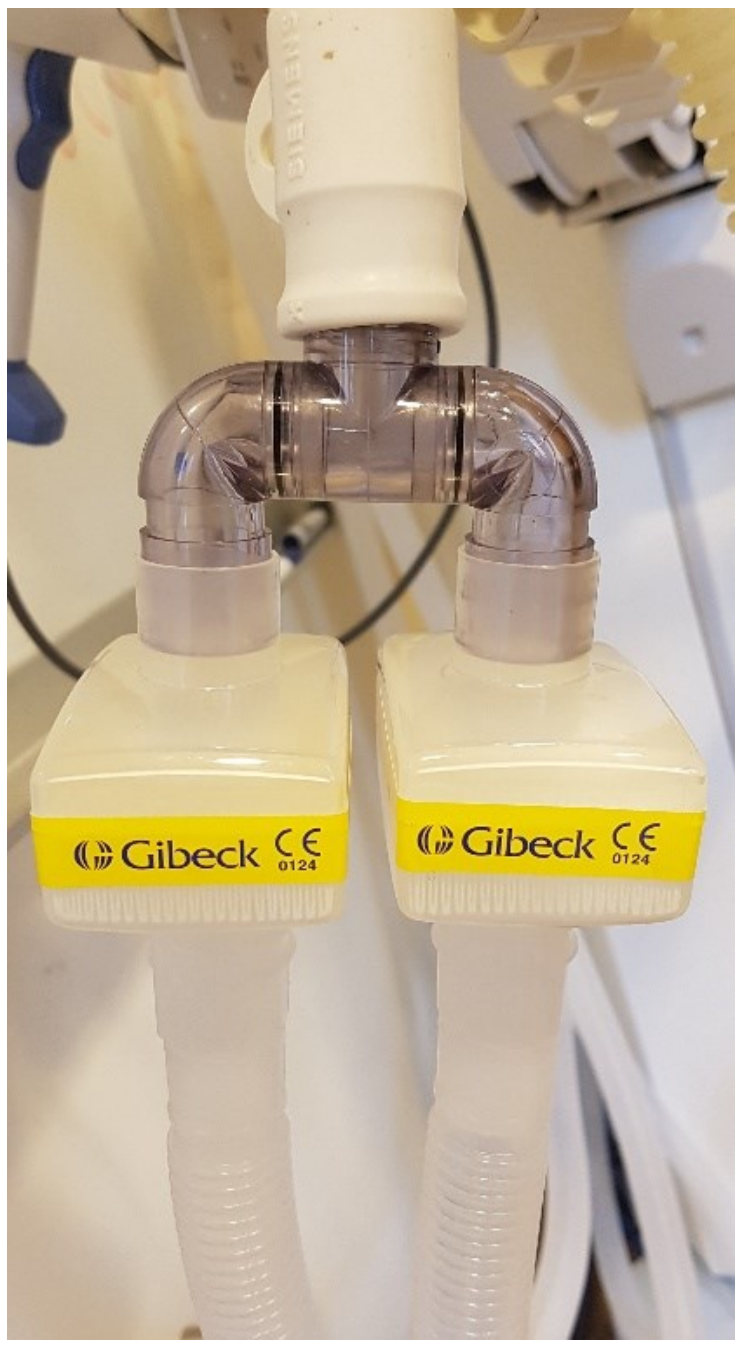

Figure 2. Splitting of the expiratory limb with a T-connector and two microbial filters for two patients.

## Relevant tests to date

Neyman and Irvin 2006 did a simulation study of the method on four equal test-lungs (Puritan-Bennet) that were ventilated for about six hours using pressure-controlled ventilation (peak pressure of 25 cmH20) followed by another six hours of volume-controlled ventilation (2000 mL tidal volume). The positive end-expiratory pressure (PEEP) was set to 0 cm $H_20$, the respiratory rate to 16 breaths/min for both modes, and the circuit was inspected approximately every 20 minutes visually checking the test lungs and recording the ventilator display. A relatively even distribution in tidal volume between the four lungs was recorded with 471±22 mL for pressure control and 507±1 mL for volume control. The authors indicate further studies for testing of efficacy and safety, but suggest a significant potential for the expanded use of a single ventilator during cases of disaster surge involving multiple casualties with respiratory failure.

Paladino *et al.* 2008 went further and tested the principle in an animal study on four 70 kg sheep ventilated for 12 hours. The method was similar to Neyman and Babcock Irvin, but with microbial filters added on each expiratory limb. Arterial pH, $pCO_2$, $pO_2$, heart rate and blood pressure was monitored during the course of the experiment. Although there were issues with $CO_2$ retention and poor oxygenation due to positioning, and a case of hypoxia due to adverse reaction to induction agents, the authors report successful oxygenation and ventilation of all animals for 12 hours while keeping them hemodynamically stable.

Branson *et al.* 2012 did a study similar to Neyman and Babcock Irvin, but with test lungs having variable airway resistance and compliance (the inverse of stiffness). In this way, they could study how the method would work if patients with different lung characteristics would be connected to the same ventilator. They found that the ventilation was distributed quite evenly for lungs with equal resistance and compliance, but found large differences in the ventilation between lungs (from 257 to 621 mL tidal volume) having different compliance. The authors recommend avoiding the use of this method, given the potential hazardous and untoward complications.

Lately, Chatburn et al did a simulation study using breathing simulators with evidence-based lung models having differences in lung resistance and compliance, combining six pairs of simulated patients. Their experiments confirmed the potential for markedly different ventilation and oxygenation for patients with uneven respiratory system impedances during multiplex ventilation. More recently, several innovative solutions have been presented to deal with the known issues of the technique, such as implementation of individual flow restrictors or pressure relief valves [Clarke et al 2020, Han et al 2020, Hermann et al 2020].

The only study done on humans to our knowledge is a test on two awake volunteers ventilated through facemasks for ten minutes, connected to the same ventilator (Draeger Evita XL) by Y-connectors splitting the inspiratory and expiratory limbs (Smith and Brown 2009). Ventilation was given by pressure-control with an inspiratory pressure of 30 cm $H_20$, PEEP at 2 cm $H_20$, 18 breaths/min while inspiratory and endtidal $CO_2$ was monitored at the masks. After ten minutes, both subjects' endtidal $CO_2$ were acceptable (4.7 and 5.7 kPa) and well tolerated.

Table 1. Published results on using one ventilator for multiple patients (does not include more advanced solutions for ventilator sharing).

| **Study** | **Model** | **Reported results** |
|---|---|---|
| Neyman and Babcock Irvin 2006 | 4 equal test lungs ventilated approximately 6 hours | Evenly distributed ventilation among all test lungs |
| Paladino *et al.* 2008 | 4 sheeps ventilated for 12 hours | Sufficient ventilation, oxygenation and hemodynamic stability throughout the experiment |
| Smith and Brown 2009 | 2 human volunteers ventilated for 10 minutes | Acceptable $CO_2$ levels after 10 minutes of ventilation |
| Branson *et al.* 2012 | 4 test lungs with different combinations of resistance and compliance | Even distribution of ventilation for equal lungs, large differences in ventilation between lungs with different compliance |
| Chatburn et al 2020 | Breathing simulators with evidence-based lung models, six pairs | Confirmation of the potential for markedly different ventilation and oxygenation for patients with uneven respiratory system impedances during multiplex ventilation. |

## Summary and discussion of previous studies

The combined data from studies related to the technique is scarce and not very relevant to ARDS, making it difficult to draw any conclusion on the suitability of the method (see table 1 for an overview). The method has been discussed and faced criticism among experts, cautioning against usage.

These tests have shown that the technique may work technically under given circumstances, but with the lack of relevant tests and the potential hazards, the method has received strong criticism since first published. Branson published comments on both the studies from Neyman and Babcock Irvin and the Paladino study in letters to the Editors in Acad. Em. Med and Resuscitation (Branson 2006 and 2008), pointing to weaknesses of the studies and raising concerns on patient safety. Concerns are raised with the animal study (Paladino *et al.* 2008) for being too optimistic with the animals having episodes of hypoxia and hypercapnia despite having no deliberate pulmonary pathology, and less relevant as the animals had well-functioning lungs.

The method has been commented by an expert group on critical care [Lewis *et al.* 2007] where the issues of dynamic airflow resistance and compliance throughout the duration of ventilation was raised along with the need for pharmacological paralyzation, seeing more potential utility for the technique on patients with normal lungs and thereby freeing additional ventilators for patients with high resistance or low compliance. Recently, protocols have been developed for implementation of the method around the world, and at the same time, the Society of Critical Care Medicine has published a *Consensus Statement on Multiple Patients Per Ventilator* (March 26 2020), advising against the use of the technique (www.sccm.org/Disaster/Joint-Statement-on-Multiple-Patients-Per-Ventilato). There is wide agreement that this technique is only a last resort solution in case of equipment shortage [Tonetti et al 2020, Pearson et al 2020].

Due to the lack of data on the feasibility of the technique, we conducted a study with a series of tests, extending the relevant cases of lung characteristics and ventilator settings tested previously. The aim of this study was to learn more about the feasibility and issues with the technique and provide more data from relevant experiments to support the evaluation of the concept from a technical standpoint.

# Methods

Two types of test lungs, either a pair of Michigan Instruments (MI) adult lung simulators or a pair of IMT Analytics Smartlung Adult were connected to one ventilator (Löwenstein Medical Leon Plus anesthetic machine) using T-connectors as shown in figure 3. The Michigan Instruments test lungs were spring-based with 2L capacity while the IMT test lungs were 1000 mL bags with a capacity of 600 mL. One of the test lungs (Lung A) was set to a constant compliance at approximately 30 mL/mbar, while the other test lung was set to variable levels of compliance as shown in Table 2. The true compliance of the test lungs were measured by the ventilator, and differed slightly for the given values for the IMT test lungs (28 mL/mbar for Lung A at 25 mL/mbar and 7, 14, 20 and 29 mL/mbar at the four adjustable levels of Lung B of 10, 15, 20 and 30 mL/mbar). Both IMT test lungs had adjustable airway resistance at the connection into the lung, possible to adjust to 5, 20, 50 and 200 mbar/L/s. For monitoring of individual tidal volumes and pressures into the lungs, an IMT Analytics type PF300 flow and pressure analyzer was connected between each lung and the assigned inspiratory and expiratory tubes as shown in figure 3.

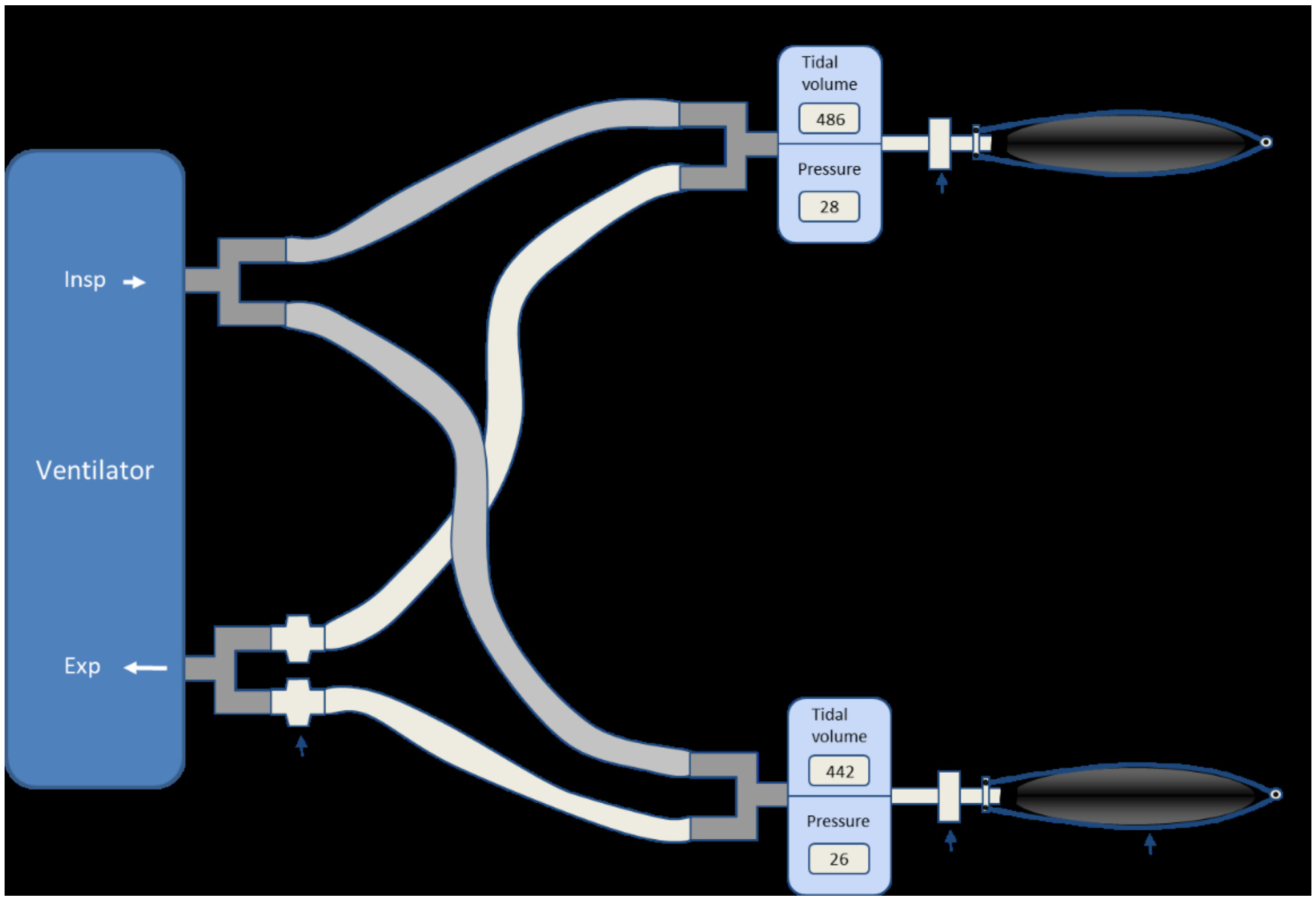

Figure 3. Setup for the testing of two lungs sharing one ventilator with monitoring of tidal volumes and pressure into both lungs (the IMT test lungs are shown in the drawing)

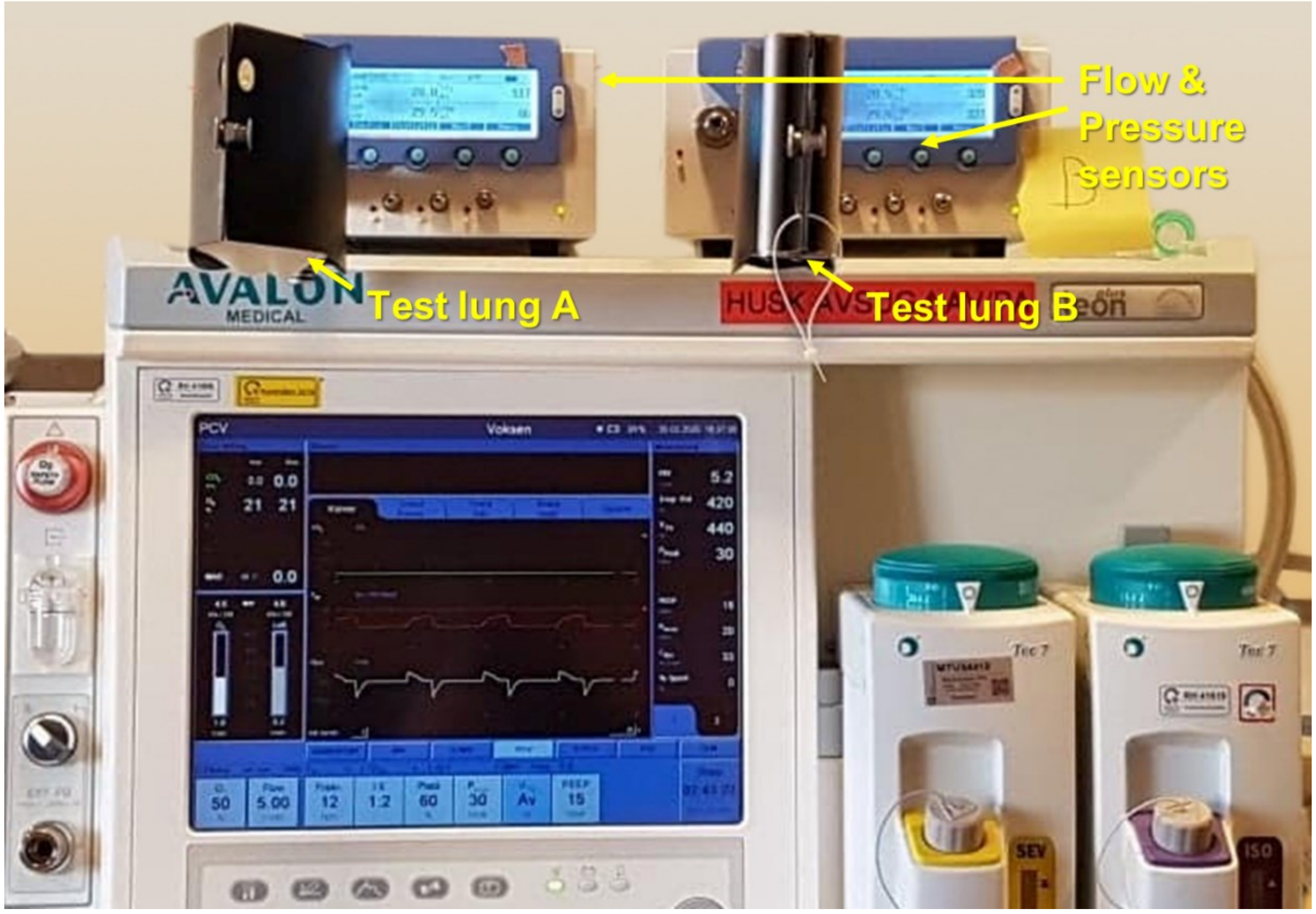

Figure 4. Test lungs, flow and pressure sensors and the anesthetic machine used in the study.

The aim was to investigate this simple ventilator sharing setup in how the distribution in ventilation between two lungs depends on differences in compliance, and how this dependency relates to the airway resistance, mode of ventilation (pressure or volume control), levels of peak inspiratory pressure (PIP) and the extrinsic positive end-expiratory pressure (PEEP). Table 2 shows all settings used in the tests. For volume-controlled ventilation, we used a tidal volume of 900 mL to obtain 450 mL in both lungs upon balanced ventilation. An earlier test with the same setup showed that the distribution between the two test lungs was independent of the tidal volume setting (from 900 to 1500 mL), hence we used only one setting at 900 mL for the volume controlled tests.  As default, a PEEP of 6 mbar was used unless mentioned otherwise.

Table 2. Settings for the ventilator and test lungs used in the study.

| Setting | Levels | | | | |
|---|---|---|---|---|---|
| Compliance of constant MI Lung A [mL/mbar] | 30 | | | | |
| Compliance of variable MI Lung B [mL/mbar] | 50 | 30 | 20 | 15 | 10 |
| Compliance of constant IMT Lung A [mL/mbar] | 28 | | | | |
| Compliance of variable IMT Lung B [mL/mbar] | 29 | 20 | 14 | 7 | |
| Airway resistance of IMT lungs [mbar/L/s] | 5 | 20 | 50 | | |
| Ventilator mode | Volume controlled | Pressure controlled | | | |
| Pressure limitation (volume-control) [mbar] | 30 | No limit (set to 80) | | | |
| PIP (pressure-control) [mbar] | 20 | 30 | | | |
| Positive end-expiratory pressure (PEEP) [mbar] | 6 | 10 | 15 | | |
| Tidal volume (volume-control) [mL] | 900 (450x2) | | | | |
| I:E ratio | 1:2 | | | | |
| Breaths / minute | 12 | | | | |
| O2 concentration | 50% | | | | |

# Results

## Modes of ventilation

Figure 5 shows how the distribution in ventilation between the two MI test lungs was dependent on the differences in compliance between them using both volume- and pressure-controlled modes of ventilation. Expectedly, the tidal volumes were equal upon equal compliances ($C_A = C_B = 30$ mL/mbar), and Lung B received lower parts of the total tidal volume as its compliance decreased. The change in Lung B's compliance also caused the ventilation of Lung A to change when in volume-controlled mode (Figure 5a), but not in pressure-controlled mode (Figure 5b and c). In volume control and at $C_B = 15$, as an example where one lung was set to half the compliance of the other, the less compliant lung received about 30% less volume while the more compliant lung received a similarly large increase in volume compared to when the compliances were equal ($C_B = 30$). In pressure controlled mode, this reduction in delivered volume to lung B at $C_B = 15$ was about 50%, but with no change to Lung A. The ratio of tidal volumes delivered between the two test lungs was approximately equal to the ratio of compliance between the lungs, independent on whether volume control or pressure control with a PIP of 20 or 30 mbar was used.

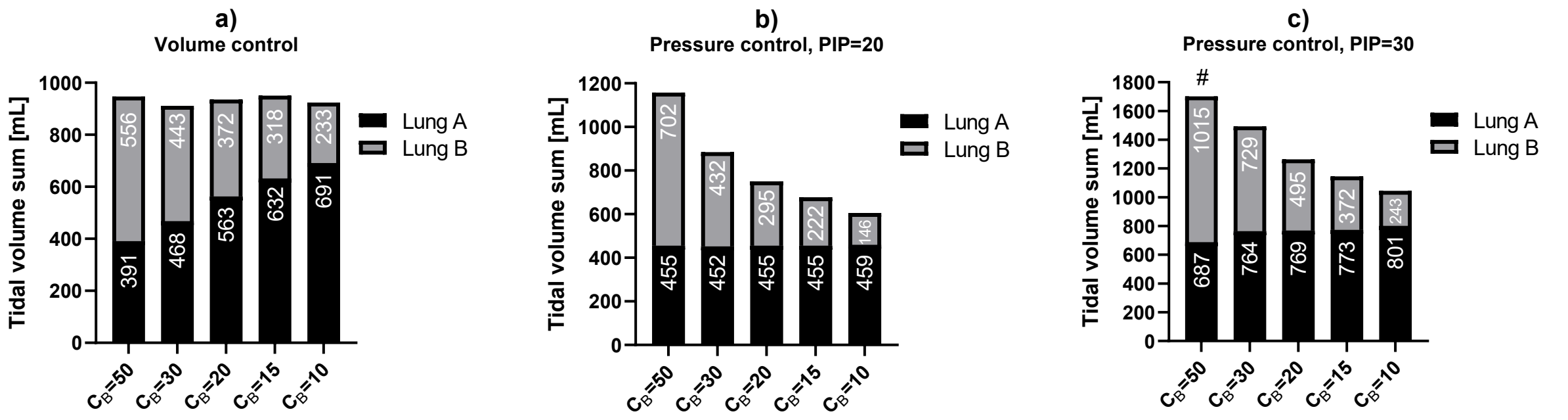

Figure 5. Delivered tidal volumes to the two MI test lungs using volume control (a), pressure control with PIP=20 mbar (b) and pressure control with PIP=30 mbar (c). Testlung A has a constant compliance of 30 mL/mbar while the compliance of testlung B ($C_B$) is changed from 50 to 10 mL/mbar. A PEEP of 6 mbar was used in these measurements. # indicates that the ventilator was not able to completely obtain the set inspiratory pressure of 30 mbar (approximately four mbar lower) due to the large MI test lung volume.

Figure 6 shows how the distribution in ventilation when the IMT test lungs were used. As the pressure limit alarms were triggered at the 30 mbar limit in volume controlled mode (figure 6a), volume-controlled ventilation without pressure limitation is also presented in figure 6b. Based on figure 6b-d, the same effect of changes in ventilation to both lungs upon changes in $C_B$ is seen in also in volume-controlled ventilation of these test lungs, along with the unchanged ventilation of lung A when in pressure-controlled mode. For these test lungs, the ratio of tidal volumes could differ from the compliance ratios, seen in particular for $C_B$=20 in pressure controlled mode.

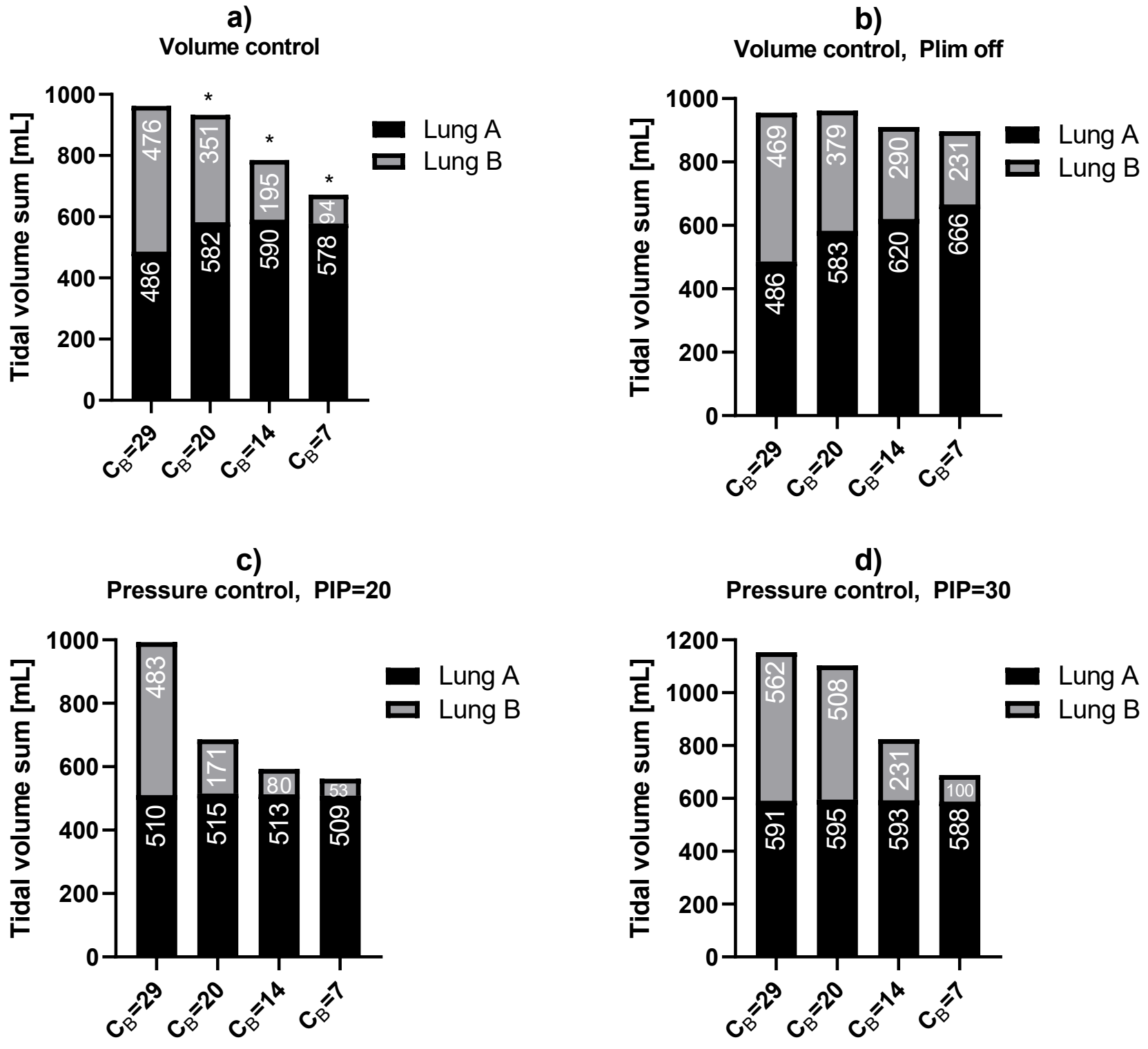

Figure 6. Delivered tidal volumes to the two IMT test lungs using volume control (a-b) and pressure control with PIP set to 20 mbar (c) and 30 mbar (d). Tidal volumes delivered with a pressure limitation at 30 mbar is shown in a) and in b) without pressure limitation. * indicates that the pressure limit was reached during the ventilation. A PEEP of 6 mbar was used in these measurements.

## The effect of PEEP setting

Figure 7 shows the distribution in ventilation between two MI test lungs when PEEP was increased to 10 and 15 mbar for volume controlled and pressure controlled ventilation. This distribution remained similar to the original setting with a PEEP at 6 mbar (figure 5) for both ventilation modes. Expectedly, the tidal volumes decreased for high PEEP due to lower driving pressures, but the ratios between the tidal volumes measured in the two lungs remained similar, independent on PEEP or PIP level. For the IMT test lungs however, the PEEP level affected the distribution in volumes between the lungs. A particularly noteworthy case was for $C_B$=20, where a PEEP level of 10 mbar would seem to ventilate the two lungs more evenly than for a PEEP at 6, but when the PEEP was increased to 15, Lung A (the more compliant of the two) lost tidal volume while lung B remained at the same level. In this case, the PEEP at 15 mbar had caused Lung A to remain highly inflated after expiration leaving little capacity for additional ventilation, while lung B was empty after expiration allowing more air to enter from the driving pressure. This effect of PEEP was present in both volume and pressure controlled ventilation.

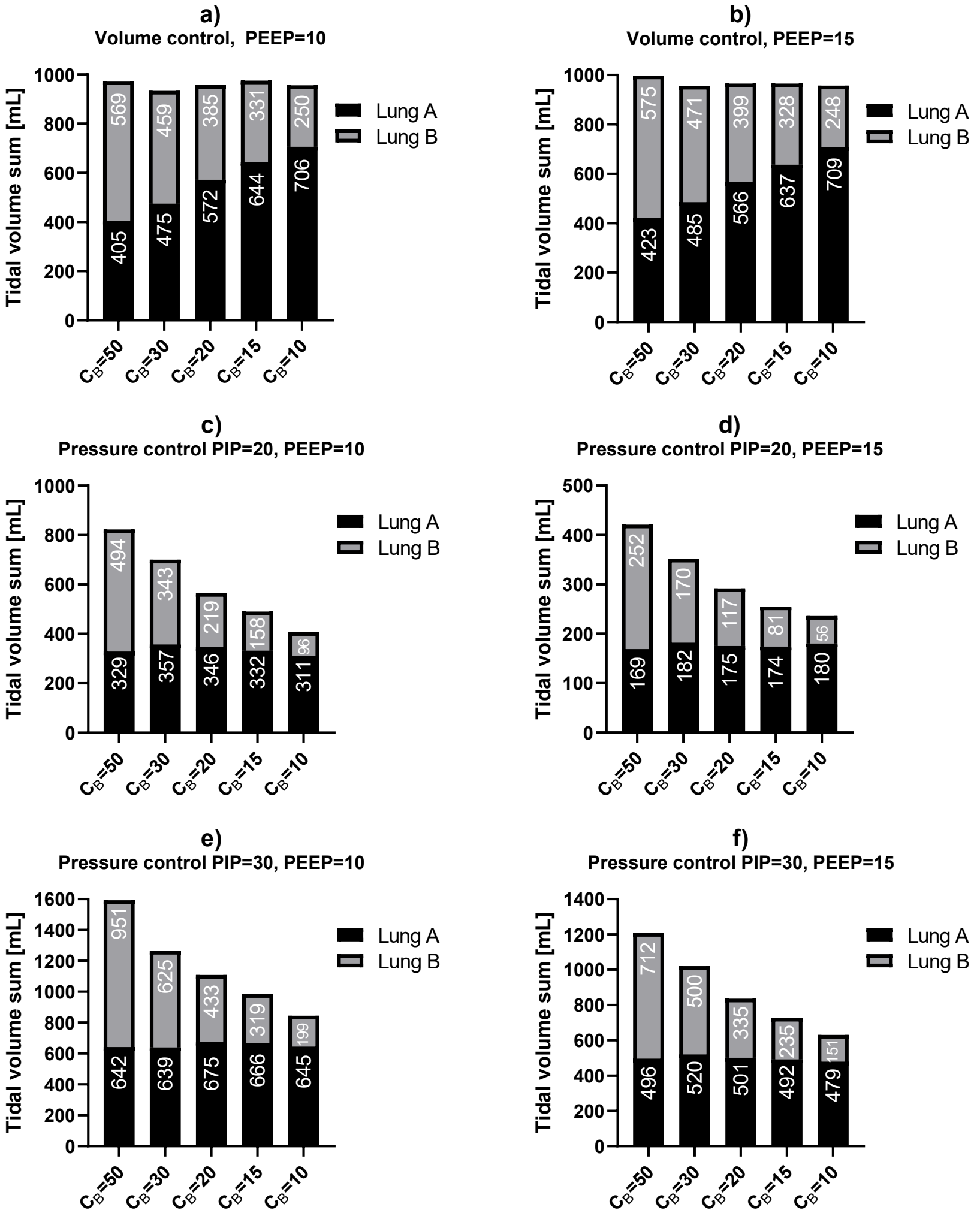

Figure 7. Delivered tidal volumes to the two MI test lungs for higher levels of PEEP, shown for volume control (a-b), pressure control with PIP=20 mbar (c-d) and pressure control with PIP=30 mbar (e-f). Testlung A has a constant compliance of 30 mL/mbar while the compliance of testlung B ($C_B$) is changed from 50 to 10 mL/mbar.

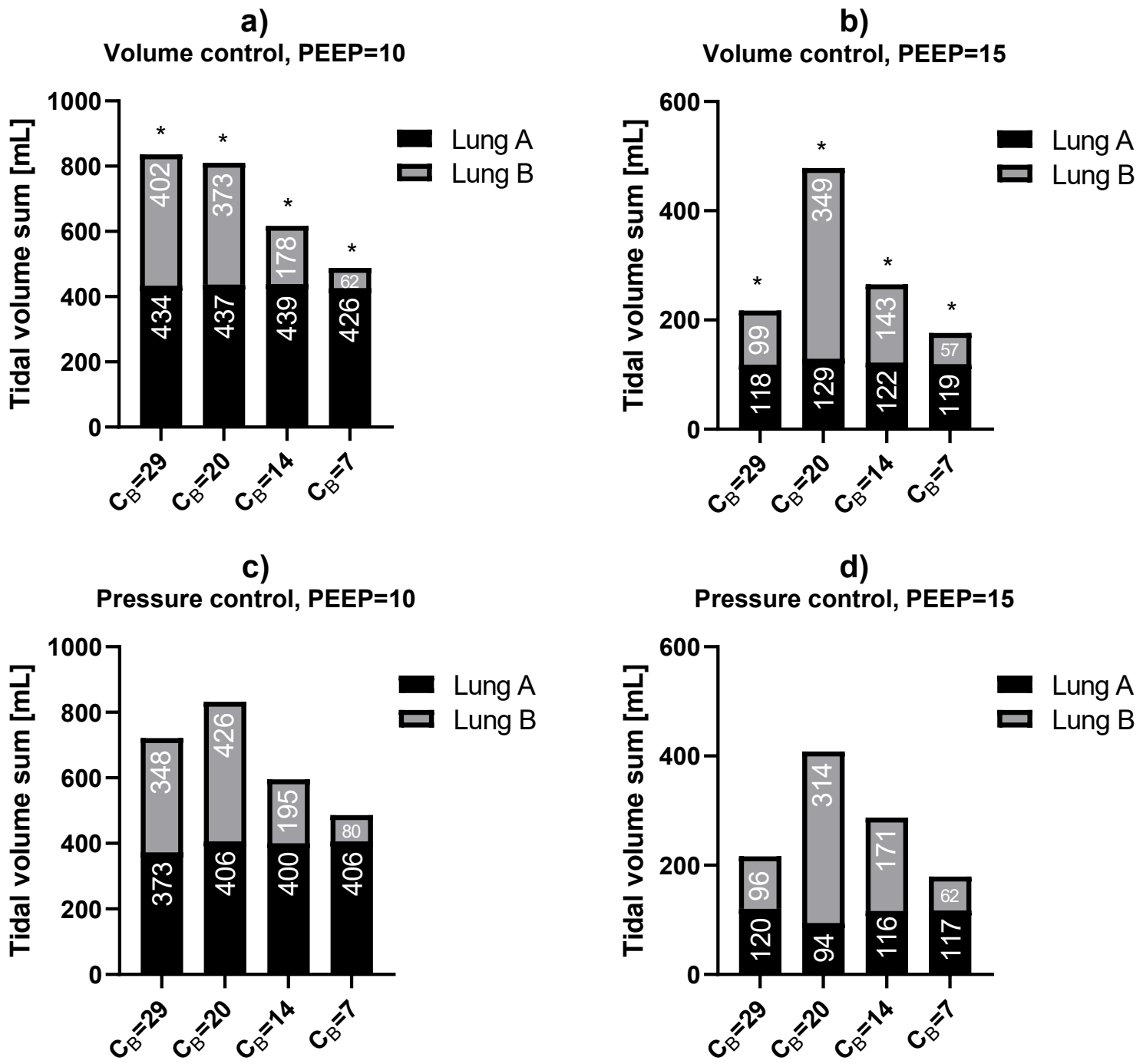

Figure 8. Delivered tidal volumes to the two IMT test lungs for higher levels of PEEP, shown for volume control (a-b) and pressure control with (c-d) and pressure control with PIP=30 mbar. Testlung A has a constant compliance of 30 mL/mbar while the compliance of testlung B ($C_B$) is changed from 30 to 7 mL/mbar. * indicates that the pressure limit was reached during the ventilation.

## The effect of airway resistance

For the IMT test lungs, we also changed the airway resistances to both lungs in different combinations, and this effect on the distribution of tidal volumes is presented in figures 9 and 10 for volume-controlled and pressure controlled modes respectively. A comparison between figure 9a (with $R_A$= $R_B$=20 mbar/L/s) and 6a (with $R_A$= $R_B$=5 mbar/L/s) shows that the volume distribution between the test lungs was unaffected by increasing the resistance equally for both lungs, independent of the compliance in Lung B. Increasing only the resistance into Lung B gave a slight reduction in the tidal volume of Lung B, and correspondingly a slight increase in tidal volume of Lung B was measured upon increasing only the resistance into Lung A. This change was small (for example 28 mL at $C_B$=20 when increasing $R_A$ to 20) although the resistance was quadrupled, and there were no significant changes in pressure. This effect of differences in airway resistance on the tidal volume was low in both volume and pressure controlled ventilation modes. A tenfold increase in the airway resistance into lung A caused a considerable decrease in the tidal volume delivered to this lung as shown in figures 9d and 10d. At the same time and only for volume controlled mode, the tidal volumes delivered to lung B increased considerably (for $C_B$=30 and 20) as a consequence of this large increase in airway resistance into lung A. In this case, the restriction ($R_A$=50) had caused the pressure in to lung A to drop to around 15 mbar.

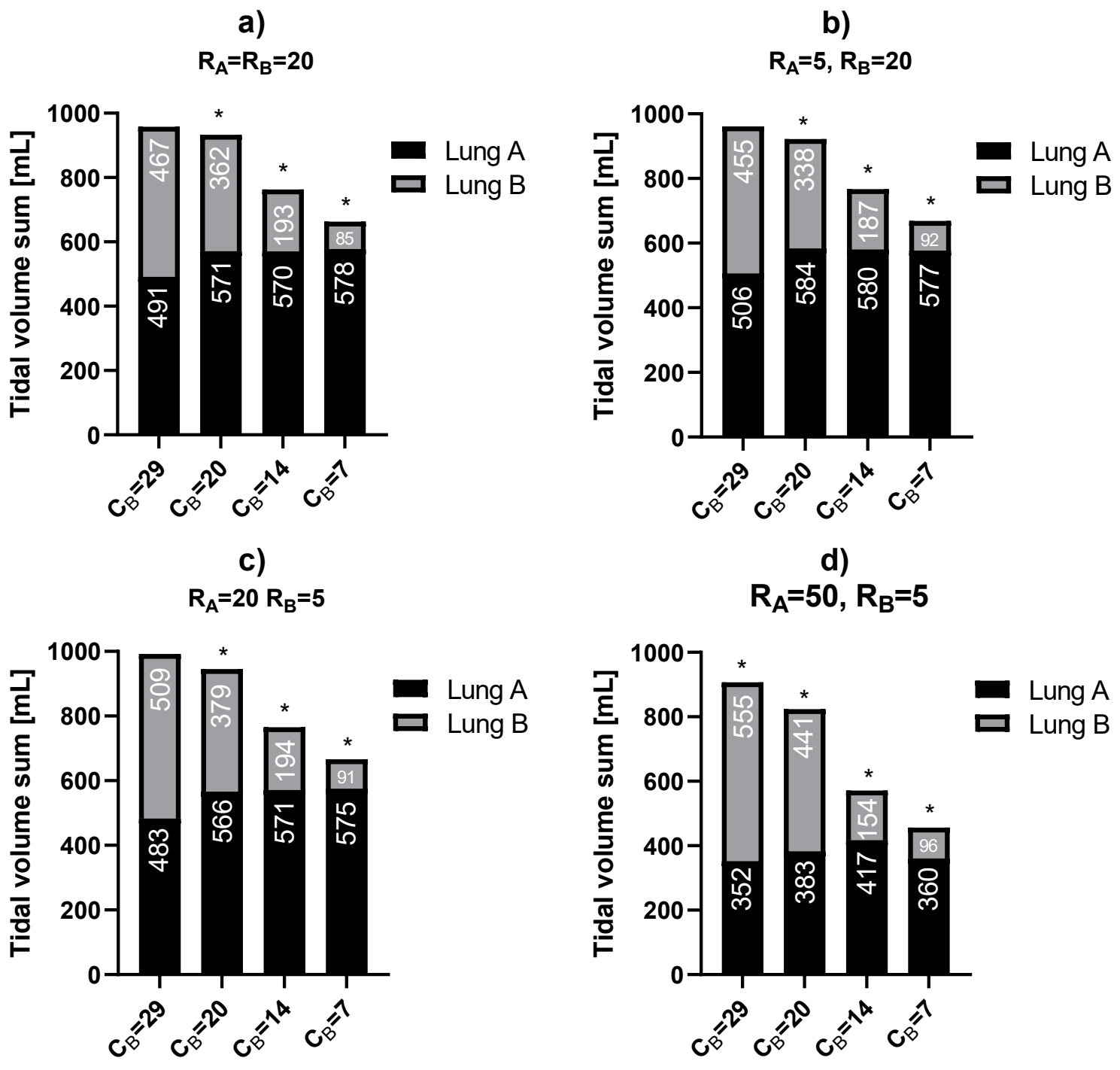

Figure 9. Delivered tidal volumes to the two IMT test lungs for different combinations of airway resistance to lung A ($R_A$) and to lung B ($R_B$), ventilated in volume control mode. A PEEP of 6 mbar was used in these measurements. * indicates that the pressure limit was reached during the ventilation.

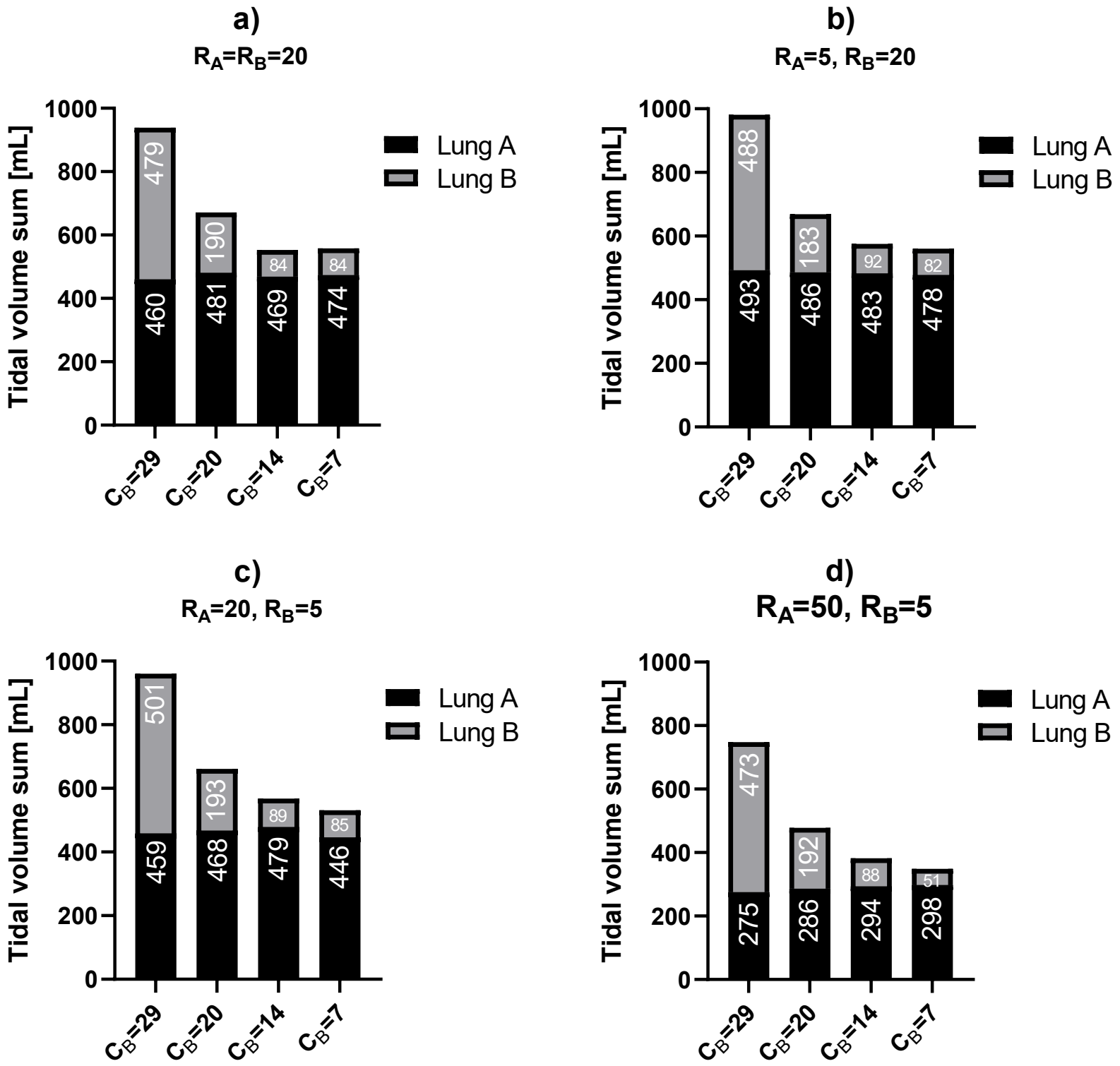

Figure 10. Delivered tidal volumes to the two IMT test lungs for different combinations of airway resistance to lung A ($R_A$) and to lung B ($R_B$), ventilated in pressure control mode. A PIP of 20 mbar and a PEEP of 6 mbar was used for these measurements.

## Discussion

The results clearly show how differences in compliance could cause large discrepancies in ventilation between the two test lungs and that this discrepancy was large already when the stiffer lung had a 30% lower compliance than the other lung. This confirms that connecting patients with such differences in compliance by this technique may cause very different delivery of tidal volumes from the same setting on the ventilator. In addition to insufficient ventilation of the stiffer lung, the problem of potentially causing volutrauma in the more compliant lung arises, in particular for volume-controlled ventilation where reduced compliance in one lung directly increases the volume delivered to the other lung. As demonstrated in the results, pressure-controlled ventilation kept the ventilation to the constant lung (A) stable while the compliance of the other lung was changed (figure 5b-c) which is an advantage of this mode. However, compensating for low ventilation of the stiffer lung by increasing PIP would still cause unintended elevated inflation of the more compliant lung (comparing figure 5b and 5c). Only two test lungs were connected in this study, and these issues would not be any less problematic when connecting four patients to the same ventilator. Two randomly paired patients with ARDS could have relatively large differences in compliance, as the baseline variation between patients can be large (Amato et al. 1998, Gattinoni et al. 2006, Sundareasn et al. 2011), with a coefficient of variation of about 20% to 40% from these studies. Perhaps even more important is the intraindividual changes in lung condition during the course of treatment and changes in lung impedance due to regular events such as changing the position of the patient, sudden obstruction due to secretion, coughing, or breathing in non-paralyzed patients.

Our results are largely in agreement with Branson *et al.* 2012, who also found a low effect of airway resistance differences compared to compliance differences in the volume distribution between test lungs. Branson *et al.* 2012 also reported that the discrepancies in delivered tidal volumes was exacerbated for pressure-controlled ventilation (using a PIP of 15 cm $H_20$), where we measured equal discrepancies for the two modes when using the MI test lungs (figure 11a). We also measured a higher discrepancy for pressure-controlled with PIP=20 and lower discrepancies with PIP=30 for the IMT test lungs, probably due to these particular test lung characteristics.

The extrinsic PEEP was set to 6 mbar as default for all tests until the PEEP was increased to 10 and 15 mbar while studying the effect on the ventilation distribution. For the MI test lungs, there was no effect on the ventilation distribution from the changes in PEEP, but for the IMT test lungs the PEEP level interacted with the distribution of tidal volumes. Going from 6 to 10 mbar PEEP, the tidal volumes of the two lungs became more equal, especially notable when $C_B$=20 (figure 8). Our next finding was that further increases in PEEP could greatly influence the tidal volumes of the two test lungs and in unintended ways. The PEEP setting is used in treatment to sustain a positive pressure after the end of expiration to prevent alveolar collapse, but this effect could be quite different when several lungs are connected to the same inspiratory and expiratory valves. As shown in figure 8, the distribution in tidal volumes was strongly dependent on the PEEP, and at $C_B$=20 Lung B obtained about three times the tidal volume of Lung A although Lung A was more compliant. In this case, Lung A was largely inflated after expiration with less inspiratory capacity left, while lung B was empty after expiration. The "two-balloon effect" (Merritt and Weinhaus 1978) could possibly have contributed to this volume discrepancy, and the inclusion of one-way valves could possibly help. This issue was only present for the IMT test lungs (behaving more like a balloon and having a lower capacity than the MI lungs), but could indicate an additional concern in ventilating lungs with different compliances using higher PEEP, also supporting a need for the possibility to adjust PEEP individually if ventilators would be shared.

According to theory, the distribution in ventilated volumes between the lungs should simply be equal to the ratio of dynamic compliances between them, assuming equal pressures into the lungs and equal lung capacities. Figure 11 shows the relation between compliance ratio between the test lungs and the fraction of the total tidal volume delivered to test lung B (having variable compliance) presented for all the experimental cases measured in this study. In summary, none of these experimental cases could reliably provide balanced ventilation to test lungs with unequal compliance. Clearly, all measurements with the MI test lungs had a good agreement with theoretical expectation, independent on modes of ventilation, level of PIP or level of PEEP. For the experiments with IMT test lungs, there were several cases deviating from the theoretical line. The cases when PEEP was set to 15 mbar, the cases with large airway resistance into lung A and for pressure-controlled ventilation with a PIP at 30 mbar all had an elevated fraction of volume delivered to lung B. On the other hand, all cases of pressure control with a PIP at 20 mbar had a reduced fraction of volume delivered to lung B. With respect to splitting of a ventilator, the line should ideally lie around a 50% fraction independent on the compliance ratio. In a couple of cases (pressure controlled ventilation, PIP=30 mbar and PEEP of either 6 or 10 mbar), the line is close to 50% from 70% to 100% compliance ratios. However, this is most likely due to the characteristics of these particular test lungs (such as their inspiratory capacities being reached), and would not be a reliable setting for this purpose. At the same time, limited inspiratory pressures (in particular the plateau pressure < 30 cmH20) is important to prevent ventilator-induced lung injury in ARDS [Alhazzani et al 2020].

The large increase in airway resistance into lung A ($R_A$=50) shifted the volume fraction closer towards 50% when the lungs had different compliance (70% compliance ratio), an effect that is relevant to the application of flow restrictors that has been suggested as a means to balance tidal volumes in split ventilation [Clarke et al 2020]. This involves connecting a flow restrictor valve into one or both of the inspiratory tubes for constriction of the flow to the more compliant lung in order to compensate for tidal volume differences. Based on our test, there was little effect on tidal volumes from a four-fold increase in constriction, and a ten-fold constriction (from 5 to 50 mbar/L/s) was necessary to have a balancing effect on the tidal volumes for the IMT test lungs with $C_A$=28 and $C_B$=20. With such a strong constriction, it could be difficult to tune the flow accurately within the remaining cross-sectional area of the tube unless the valve control is very precise. In addition, such constriction and increase of airway resistances alters to the inspiratory time constant (resistance multiplied by compliance). We believe that pressure-reduction valves would be more suitable for this purpose, providing a more linear control of individual inspiratory pressures and tidal volumes.

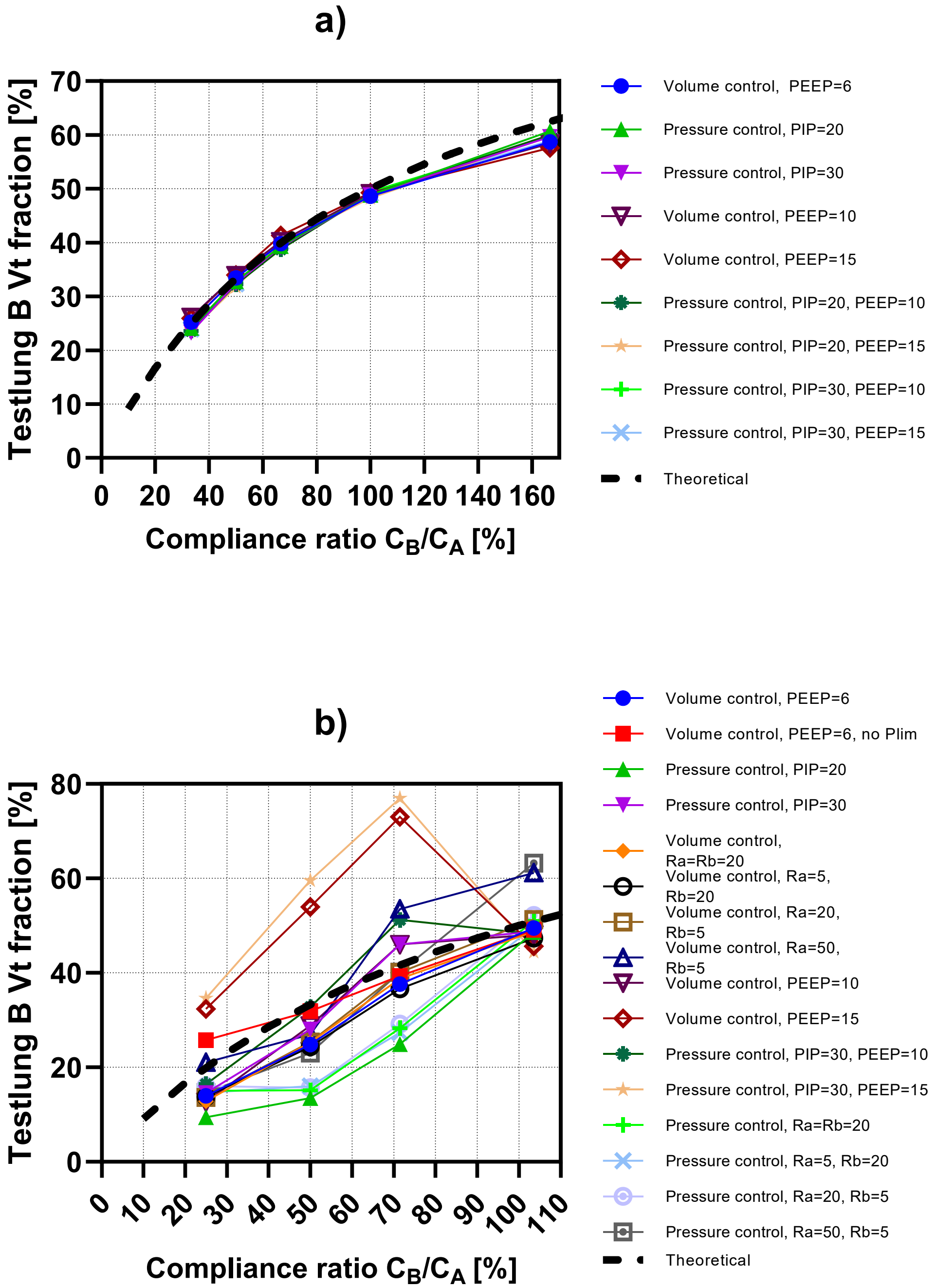

Figure 11. The fraction of the total tidal volume delivered to the test lung with variable compliance (lung B) plotted as a function of the ratio of compliances between the test lungs. All experimental cases are plotted for the MI test lungs in a) and for the IMT test lungs in b). Note that a) includes one measurement with higher compliance ratio (167%) and that b) also includes tests with different airway resistances.

It is important to stress that experimenting on test lungs have obvious limitations and is far from the real situation with ventilation of real pathological lungs over many days with additional concerns such as contamination, coughs and voluntary breathing triggering the ventilator, and the extrapolation of these experiments to the treatment of real lungs is limited. In addition, there are ethical concerns, as discussed in Pearson et al 2020. Although two different types of test lungs were used, both will likely have different characteristics from lungs in ARDS patients, including the levels of compliance and the pressure-volume characteristics. The IMT test lungs had adjustable compliance from 10 to 30 mL/mbar (in reality 7-29 mL/mbar) and a capacity of 600mL, which is far less than expected for healthy lungs and also low for ARDS lungs. The IMT test lungs may be more relevant for the case of the "baby lung" associated with ARDS [Gattinoni and Pesenti 2005]. The discrepancy in results between the two types of test lungs also underscores the importance of properly relevant test lungs and their characteristics when testing new methods in simulations. The tests in this study was done using only one particular ventilator, a part of an anesthetic machine representing a typical ventilator, and the ventilator behavior when connecting multiple circuits may differ between models.

Anyhow, our main goal was to investigate the distribution of ventilation for lungs with relative differences in compliance. At the same time, this should be taken into consideration when interpreting our results on patients with other compliance levels. The observed effects of PEEP in test lung with different compliance will likely be of clinical relevance in patients with ARDS as personalized, elevated PEEP often is indicated. An issue with differences in the effect of PEEP due to even modest differences in compliance between patients was also reported in Chatburn et al 2020. In patients with severe lung failure due to ARDS the optimal combination between pressure, flow and PEEP is sought. Individualized therapy will not be possible when one ventilator is shared between two or more patients unless substantial modifications are implemented. Sharing one ventilator on two or more patients will complicate monitoring of all included patients and the competence and clinical skills of the critical care nurse or physician will be crucial. The ventilator setting is individualized based on measurements of flow, pressure, and end-tidal $CO_2$. In a patient safety perspective, a simple pressure controlled ventilator mode with moderate PEEP in patients with comparable respiration physiology is indicated.

## Conclusion

Based on the test in this study of the simple ventilator splitting technique [Neyman G and Irvin CB 2006] including different test lungs, ventilation modes, PIP levels, PEEP levels and combinations of airway resistances, we did not find any reliable setting, adjustment or simple measure to provide balanced ventilation of lungs with different compliance. A more advanced modification is indicated that includes in-line pressure-reduction valves at the inspiratory limbs for individualized inspiratory pressure, individual adjustment of PEEP at the expiratory limbs (potentially also by in-line pressure reduction valves), sensors for individual monitoring of tidal volumes and pressures in addition to the splitters and filters used in the simple setup. Such a solution has lately been in development, investigated and tested by an international working group on differential multiventilation (differentialmultivent.org).

## Acknowledgements

The authors would like to thank Professor Richard Branson from the University of Cincinnati and Professor Erwan L'Her from Université de Bretagne Occidentale for sharing of knowledge, experiences and helpful discussions. The authors would also like to thank the international differential multiventilation collaboration initiative coordinated by Hannah Pinson from the University of Vrije, Belgium (differentialmultivent.org).

## Dataset

The measured data are available upon request.